\documentclass{appolb}
\usepackage{graphicx}

\begin{document}
\title{Tensor Glueball in a Top-Down Holographic Approach to QCD%
\thanks{Presented at ``Excited QCD 2015\textquotedblright, Tatranska Lomnica, Slovakia, March 8-14, 2015.}%
}
\author{Denis Parganlija
\address{Institut f\"{u}r Theoretische Physik, Technische Universit\"{a}t Wien, Wiedner Hauptstr.\ 8-10, 1040 Vienna, Austria}
\\
}
\maketitle
\begin{abstract}
Properties of the tensor glueball are discussed in the Witten-Sakai-Sugimoto Model, a top-down holographic approach to the non-perturbative region 
of Quantum Chromodynamics (QCD).
\end{abstract}
  
\section{Introduction} \label{Section1}
The existence of glueballs -- bound states of gluons, the gauge bosons of Quantum Chromodynamics -- is naturally expected due to the non-Abelian nature of
the theory \cite{Salomone:1980sp}.
As a matter of principle glueballs can possess various quantum numbers, the most important of which are the total spin $J$, 
parity $P$ and charge conjugation $C$. In practice, however, their experimental identification
is often problematic given the possible overlap with quark bound states carrying the same quantum numbers \cite{Mathieu}.
Nonetheless, simulations in lattice QCD suggest the existence of a glueball spectrum 
\cite{Morningstar:1999rf}.
\\\\
Glueballs are noteworthy for at least two reasons. The Brout-Englert-Higgs mechanism, responsible for non-vanishing current 
quark masses, plays no role in the mass generation of glueballs that is a consequence of only the strong interaction. 
Additionally, glueballs possess integer spin and are thus classified as mesons whose spectrum without glueballs would 
be incomplete.
\\\\
According to lattice-QCD simulations, the lightest tensor glueball has a mass
between 2.3 GeV and 2.6 GeV \cite{Morningstar:1999rf}.
Listings of the Particle Data Group (PDG \cite{PDG}) contain four tensor states around 2 GeV termed as established:
$f_2(1950)$, $f_2(2010)$, $f_2(2300)$ and $f_2(2340)$; several other states require confirmation,
such as a narrow $f_J(2220)$ state that may have spin two or four.
Tensor states have been subject of various low-energy effective approaches 
\cite{Amsler:1995td} but a tensor glueball has still not been clearly identified.
\\\\
In this article a different approach is described: properties of the tensor glueball are examined by means of holographic QCD.
The core of the study is the so-called AdS/CFT correspondence -- the conjectured duality between supergravity
theory (the weak-coupling limit of string theory) in an anti-de Sitter (AdS) space and a strongly coupled conformal field theory (CFT) in one dimension less.
The full space of the supergravity theory contains a compact component (n-sphere $S^n$) such that the
total number of dimensions equals 10 or 11, depending
on whether string theory or M-theory is used. As an example, the original form of the correspondence entailed the equivalence between
the supergravity limit of type-IIB string theory, containing fermions of the same chirality, on an $AdS_5 \times S^5$ space and the large-$N_c$ limit of
an $\cal{N}$ $=4$ supersymmetric and conformal $U(N_c)$ gauge theory on 
a 4-dimensional boundary of the $AdS_5$ space \cite{Maldacena}, where $\cal{N}$ represents the number of the supersymmetry generators.\\
An application of the correspondence to QCD is only possible once the supersymmetry and the conformal symmetry are removed
from the gauge theory and in Ref.\ \cite{Witten} Witten proposed a way to this goal in the 11-dimensional M-theory where, in the low-energy limit (supergravity with an $AdS_7 \times S^4$ background),
the undesired symmetries are broken by suitable circle compactifications. Then a pure Yang-Mills theory
is obtained with the supergravity limit requiring a finite radius for the supersymmetry-breaking circle (the inverse radius determines the
value of the so-called Kaluza-Klein mass $M_{\mbox{\tiny KK}}$), and also a large 't Hooft coupling. Thus constructed holographic approaches 
to QCD derived directly from string theory are 
referred to as {\it top-down} models \cite{Klebanov2000,SaSu}. An extension is the Witten-Sakai-Sugimoto (WSS) Model;
its implications for the tensor glueball presented here are based on the detailed discussions in Ref.\ \cite{BPR}.

\section{The Witten-Sakai-Sugimoto Model and Its Implications for Tensor Glueballs}
Witten's model contained no chiral quarks; in Ref.\ \cite{SaSu} a method for their inclusion was proposed
by introducing $N_f$ (number of flavours) probe D8- and anti-D8-branes [inducing $U(N_f) \times U(N_f)$ chiral
symmetry] that extend along all dimensions of the supergravity space except for a 
(Kaluza-Klein) circle. In the simplest case, the branes and antibranes are located antipodally
with regard to the circle; however, they merge at a certain point in the bulk space
reducing the original $U(N_f) \times U(N_f)$ symmetry to its diagonal subgroup which is
interpreted as a realisation of chiral-symmetry breaking.
\\\\
Up to a Chern-Simons term, the action for D8-branes reads

\begin{equation}
S_{\mbox{\tiny D8}} = -T_{\mbox{\tiny D8}} \, \mbox{Tr} \int \mbox{d}^9 x e^{-\Phi} \sqrt{- \mbox{det}(\tilde{g}_{MN} + 2\pi \alpha ' F_{MN})} \label{D8}
\end{equation}
where $T_{\mbox{\tiny D8}} = (2\pi)^{-8} l_s^{-9}$ (and $l_s^2 = \alpha '$, with $l_s$ the string 
length), trace is taken with respect to flavour, $g_{MN}$ is the metric of the D-brane world volume, $\Phi$ is the dilaton field and $F_{MN}$ a field strength
tensor whose components are, upon dimensional reduction, identified as meson fields of interest. Since no backreaction
of the Witten-model background to D8-branes is considered, $N_f$ is fixed and
significantly smaller \cite{SaSu} than the number of colours (large-$N_c$ limit).
\\\\
The action of Eq.\ (\ref{D8}) can be expanded up to the second order in fields:
\begin{equation}
S_{\mbox{\tiny D8}}^{(2)} = - \kappa \, \mbox{Tr} \int \mbox{d}^4 x \int_{-\infty }^{\infty} \mbox{d}Z
\left[ \frac{1}{2} \, K^{-\frac{1}{3}} \, \eta^{\mu \rho} \eta^{\nu \sigma} F_{\mu \nu}F_{\rho \sigma} + M_{\mbox{\tiny KK}}^2 \, \eta^{\mu \nu}
F_{\mu Z} F_{\nu Z} \right ]
\end{equation}
where $\kappa = \lambda N_c / (216 \pi^3)$ \cite{BPR}, $\lambda = g_{\mbox{\tiny YM}}^2 N_c$ is the 't Hooft coupling (and $g_{\mbox{\tiny YM}}$
the 4-dimensional coupling),
$Z$ is essentially the holographic radial coordinate (and $K = 1+Z^2$) and
$\eta^{\mu \nu}$ is the flat metric diag$(-,+,+,+)$.
There are two undetermined quantities: $M_{\mbox{\tiny KK}}$, that sets the model scale,
and the coupling $\lambda$. They are usually calculated such that the mass of the rho meson
and the pion decay constant
correspond to their physical values leading to $M_{\mbox{\tiny KK}} = 949$ MeV and $\lambda = 16.63$.
Alternative determinations of $\lambda$ and $M_{\mbox{\tiny KK}}$ shift values of decay widths
but do not alter overall conclusions regarding glueballs \cite{BPR}.
\\\\
Masses and decay widths of the scalar glueball
and its first excitation in the WSS Model have been extensively studied in Ref.\ \cite{BPR} 
(see also Ref.\ \cite{Hashimoto}). Although the mixing patterns of $\bar{q}q$ and glueball states in the spectrum of $f_0$
resonances still bear many uncertainties \cite{Parganlija:2010fz},
there are indications that the glueball ground state might be dominantly unmixed \cite{Janowski:2011gt};
the analysis of Ref.\ \cite{BPR} then prefers the $f_0(1710)$ resonance as a main candidate for the scalar glueball. This result, obtained in the chiral limit,
is supported by the estimated consequences of finite pseudoscalar masses on the scalar-glueball decay \cite{BR}, motivating exploration of the spin-two glueball
in the model.

\subsection{Mass and Decays of the Tensor Glueball in the Witten-Sakai-Sugimoto Model}
Once $M_{\mbox{\tiny KK}}$ is known, the WSS Model predicts the tensor-glueball mass to be $M_{T}$ = 1487 MeV [the mass is
the same as that of the dilaton glueball, preferentially identified as $f_0(1710)$ in Ref.\ \cite{BPR}, since the tensor and this scalar mode are associated with the same multiplet
on the gravity side of the correspondence].
\\\\
The interaction Lagrangian containing the tensor glueball and pions is \cite{BPR}:
\begin{equation}
{\cal L} = \frac{1}{2} t_1 \mbox{Tr }(T^{\mu \nu} \partial_\mu \pi \partial_\nu \pi) \label{Lagrangian}
\end{equation}
where the trace is over isospin and $t_1 = 42.195 / (\sqrt{\lambda} N_c M_{\mbox{\tiny KK}})$ \cite{BPR}.
The ensuing decay width reads (pions are massless \cite{BPR}):

\begin{equation}
\Gamma_{T \rightarrow \pi\pi} = \frac{1}{640 \pi} t_1^2 M_T^3, \label{T2pi}
\end{equation}
i.e., $\Gamma_{T \rightarrow \pi\pi} = 22$ MeV. Thus the holographic tensor glueball appears to be quite narrow in the chiral 
limit.
\\\\
It needs to be noted, however, that the result of Eq.\ (\ref{T2pi}) comes about with a tensor mass markedly lower than
the lattice-QCD result of (2.3 - 2.6) GeV, discussed in Sec.\ \ref{Section1}. Additionally, $M_T = 1487$ MeV is below the
$2\rho$ threshold whose opening can entail a significant contribution of the $4\pi$ decay channel to the total decay width
of the tensor glueball. Therefore a comparison with experimental data seems only justified if the tensor mass 
is extrapolated to the interval suggested by lattice QCD. (One also needs to keep in mind that, on the gravity side of the correspondence,
$\alpha '$ corrections
might conceivably shift the value of $M_T$ towards the lattice-QCD result.)
\\\\
The expectation is that in the holographic setup the tensor glueball does not couple to a pseudoscalar mass term. 
In that case non-vanishing pseudoscalar masses would only have a kinematic effect in $\Gamma_{T \rightarrow \pi\pi}$.
Then, in addition to raising $M_T$, the tensor decay width can be recalculated 
with $m_\pi \neq 0$. It is even possible to estimate the decay widths into kaons and etas
from $\Gamma_{T \rightarrow \pi\pi}$ by using 
flavour-symmetry factors 4:3 and 1:3, respectively, and also physical kaon and eta masses.\\
\\
There are direct couplings to vector mesons as well \cite{BPR}. It is then possible to calculate decay widths $\Gamma_{T \rightarrow K^{*} K \rightarrow 2(K \pi)}$ and $\Gamma_{T \rightarrow \phi \phi}$
by raising the vector mass in $\Gamma_{T \rightarrow \rho \rho }$ to the physical values of
$m_{K^{*}}$ and $m_\phi$, 
respectively, and using appropriate isospin factors.
All results are presented in Table \ref{Table}.
For an estimate of uncertainty note that (\textit{i}) abandoning the chiral limit may modify a tensor-vector coupling
in the model and increase the total decay width by $\simeq 5 \%$ (see Ref.\ \cite{BPR}) and (\textit{ii})
all values in Table \ref{Table} increase by $\simeq$ 30 \% when an alternative method to determine $\lambda$
is considered \cite{BPR}. 

\begin{table}[h]
\centering
\begin{tabular}{|c|c|}
\hline
Decay & Width [MeV]  \\ \hline
$T \rightarrow K^{*} K^{*} \rightarrow 2(K \pi) $ & $415 $  \\ \hline
$T \rightarrow \rho \rho \rightarrow 4\pi $ & $382 $  \\ \hline
$T \rightarrow \omega \omega \rightarrow 6\pi$ & $127$  \\ \hline
$T \rightarrow \phi \phi$ & $77$  \\ \hline
$T \rightarrow \pi \pi$ & $34$  \\ \hline
$T \rightarrow  K K $ & $29$  \\ \hline
$T \rightarrow \eta \eta$ & $6$  \\ \hline
Total & $1070$  \\ \hline
\end{tabular}%
\caption{Tensor decay widths for $M_T = 2400$ MeV and other particle masses at their respective physical values.}
\label{Table}
\end{table}
$\,$\\
The holographic tensor glueball is then extremely broad at $M_T = 2400$ MeV: its total decay width is 1070 MeV,
the main reason being large contributions from $\rho \rho$ and $K^{*}K^{*}$ threshold openings. For  $M_T = 2000$ MeV,
the total decay width decreases to 640 MeV \cite{BPR} which is at best marginally comparable to
$\Gamma = (472 \pm 18)$ MeV of the broadest
currently known tensor state, $f_2(1950)$. The existence of the pure tensor glueball may therefore 
be difficult to ascertain in experimental data; admixture of $\bar{q}q$ states may be the reason
for the relatively smaller total decay widths
of physical resonances.

\section{Summary and Outlook}

A holographic top-down approach to non-perturbative QCD -- the Witten-Sakai-Sugimoto Model -- has been presented
and its implications in the $2^{++}$ glueball channel have been discussed. Once the model coupling and scale have been
determined,
tensor decay widths
at masses close to or above 2 GeV can be calculated. Tensor decays are strongly
dominated by $K^{*} K^{*}$ and $\rho \rho$ threshold openings. The total decay width has a value above 1 GeV for glueball 
mass  $M_T = 2400$ MeV and decreases to 640 MeV at $M_T = 2000$ MeV. 
Such a large decay width would render this state difficult to observe in the data
(for example those expected from the PANDA Collaboration at FAIR \cite{PANDA}).
Results may change,
however, if there is a notable $\bar{q}q$ admixture to the glueball giving rise to the physical state (or states).
\\\\
{\bf Acknowledgments.}
I am grateful to F.~Br\"{u}nner and A.~Rebhan for collaboration and to D.~Bugg and S.~Janowski for extensive discussions.
This work is supported by the Austrian Science Fund FWF, project no.\ P26366.


\begin{thebibliography}{99}
\bibitem{Salomone:1980sp} 
A.~Salomone, J.~Schechter and T.~Tudron,
Phys.\ Rev.\ D {\bf 23}, 1143 (1981);  
C.~Rosenzweig, A.~Salomone and J.~Schechter,
Nucl.\ Phys.\ B {\bf 206}, 12 (1982)  [Erratum-ibid.\ B {\bf 207}, 546 (1982)];  
A.~A.~Migdal and M.~A.~Shifman,
Phys.\ Lett.\ B {\bf 114}, 445 (1982).  

\bibitem{Mathieu} 
  V.~Mathieu, N.~Kochelev and V.~Vento,
  Int.\ J.\ Mod.\ Phys.\ E {\bf 18}, 1 (2009).

\bibitem{Morningstar:1999rf} 
C.~J.~Morningstar and M.~J.~Peardon,
Phys.\ Rev.\ D {\bf 60}, 034509 (1999)  [hep-lat/9901004];  
  A.~Hart {\it et al.}  [UKQCD Collaboration],
Phys.\ Rev.\ D {\bf 65}, 034502 (2002)  [hep-lat/0108022];  
Y.~Chen {\it et al.},
Phys.\ Rev.\ D {\bf 73}, 014516 (2006)  [hep-lat/0510074]; 
M.~Loan, X.~Q.~Luo and Z.~H.~Luo,
Int.\ J.\ Mod.\ Phys.\ A {\bf 21}, 2905 (2006)  [hep-lat/0503038];
E.~Gregory {\it et al.},
JHEP {\bf 1210}, 170 (2012)  [arXiv:1208.1858 [hep-lat]].  

\bibitem{PDG} K.~A.~Olive \textit{et al.} (Particle Data Group), Chin.\ Phys.\ C, 
\textbf{38}, 090001 (2014).
  
\bibitem{Amsler:1995td} 
  C.~Amsler and F.~E.~Close,
Phys.\ Rev.\ D {\bf 53}, 295 (1996)  [hep-ph/9507326];  
  S.~R.~Cotanch and R.~A.~Williams,
Phys.\ Lett.\ B {\bf 621}, 269 (2005)  [nucl-th/0505074]; 
F.~Giacosa, T.~Gutsche, V.~E.~Lyubovitskij and A.~Faessler,
Phys.\ Rev.\ D {\bf 72}, 114021 (2005)  [hep-ph/0511171];  
  V.~V.~Anisovich {\it et al.},
Phys.\ Atom.\ Nucl.\  {\bf 69}, 520 (2006). 


\bibitem{Maldacena} 
  J.~M.~Maldacena,
Int.\ J.\ Theor.\ Phys.\  {\bf 38}, 1113 (1999)  [Adv.\ Theor.\ Math.\ Phys.\  {\bf 2}, 231 (1998)]  [hep-th/9711200].  


\bibitem{Witten} 
  E.~Witten,
Adv.\ Theor.\ Math.\ Phys.\  {\bf 2}, 505 (1998)  [hep-th/9803131].  

\bibitem{Klebanov2000} 
  J.~Babington {\it et al.}
  Phys.\ Rev.\ D {\bf 69}, 066007 (2004)
  [hep-th/0306018];
  M.~Kruczenski {\it et al.}
  JHEP {\bf 0405}, 041 (2004)
  [hep-th/0311270].
  
\bibitem{SaSu} 
T.~Sakai and S.~Sugimoto,
Prog.\ Theor.\ Phys.\ \textbf{113}, 843 (2005) [hep-th/0412141];
T.~Sakai and S.~Sugimoto,
Prog.\ Theor.\ Phys.\ \textbf{114}, 1083 (2005) [hep-th/0507073];
T.~Imoto {\it et al.},
  Prog.\ Theor.\ Phys.\  {\bf 124}, 263 (2010)
  [arXiv:1005.0655 [hep-th]];
  A.~Rebhan,
  arXiv:1410.8858 [hep-th].


\bibitem{BPR} 
  F.~Br\"{u}nner, D.~Parganlija and A.~Rebhan,
  Phys.\ Rev.\ D {\bf 91}, no. 10, 106002 (2015); Erratum: Phys.\ Rev.\ D {\bf 93}, no. 10, 109903 (2016)
  [arXiv:1501.07906 [hep-ph]].



\bibitem{Hashimoto} 
  K.~Hashimoto, C.-I.~Tan and S.~Terashima,
  Phys.\ Rev.\ D {\bf 77}, 086001 (2008)
  [arXiv:0709.2208 [hep-th]];
  F.~Br\"{u}nner, D.~Parganlija and A.~Rebhan,
  Acta Phys.\ Polon.\ Supp.\  {\bf 7}, no. 3, 533 (2014)
  [arXiv:1407.6914 [hep-ph]];
  F.~Br\"{u}nner, D.~Parganlija and A.~Rebhan,
  arXiv:1502.00456 [hep-ph];
  D.~Parganlija,
  arXiv:1503.00550 [hep-ph].



\bibitem{Parganlija:2010fz} 
D.~Parganlija, F.~Giacosa and D.~H.~Rischke,
Phys.\ Rev.\ D {\bf 82}, 054024 (2010)  [arXiv:1003.4934 [hep-ph]];  
D.~Parganlija, P.~Kovacs, G.~Wolf, F.~Giacosa and D.~H.~Rischke,
Phys.\ Rev.\ D {\bf 87}, 014011 (2013)  [arXiv:1208.0585 [hep-ph]].  


\bibitem{Janowski:2011gt} 
S.~Janowski, D.~Parganlija, F.~Giacosa and D.~H.~Rischke,
Phys.\ Rev.\ D {\bf 84}, 054007 (2011)  [arXiv:1103.3238 [hep-ph]];  
S.~Janowski {\it et al.}
Phys.\ Rev.\ D {\bf 90}, no. 11, 114005 (2014)  [arXiv:1408.4921 [hep-ph]].  

\bibitem{BR} 
  F.~Br\"{u}nner and A.~Rebhan,
  arXiv:1504.05815 [hep-ph].


\bibitem{PANDA} 
  M.~F.~M.~Lutz {\it et al.}  [PANDA Collaboration],
  arXiv:0903.3905 [hep-ex].

\end{thebibliography}
\end{document}